\documentclass{ws-procs961x669}            
\begin{document}

\title{Curvature-matter couplings in modified gravity: from linear models to conformally invariant theories}

\author{Francisco S. N. Lobo$^*$}

\address{Instituto de Astrofísica e Ciências do Espaço, Faculdade de Ciências da Universidade de Lisboa, Edifício C8, Campo Grande, P-1749-016 Lisbon, Portugal\\
Departamento de F\'{i}sica, Faculdade de Ci\^{e}ncias da Universidade de Lisboa, Edifício C8, Campo Grande, P-1749-016 Lisbon, Portugal
$^*$E-mail: fslobo@fc.ul.pt\\
https://ciencias.ulisboa.pt/perfil/fslobo
}

\author{Tiberiu Harko}

\address{Department of Theoretical Physics, National Institute of Physics
and Nuclear Engineering (IFIN-HH), Bucharest, 077125 Romania,\\
Department of Physics, Babes-Bolyai University, Kogalniceanu Street,
Cluj-Napoca 400084, Romania,\\
Astronomical Observatory, 19 Ciresilor Street, 400487 Cluj-Napoca, Romania\\
E-mail: tiberiu.harko@aira.astro.ro}

\begin{abstract}
In this proceeding, we review modified theories of gravity with a curvature-matter coupling between an arbitrary function of the scalar curvature and the Lagrangian density of matter. This explicit nonminimal coupling induces a non-vanishing covariant derivative of the energy-momentum tensor, that implies non-geodesic motion and consequently leads to the appearance of an extra force. Here, we explore the physical and cosmological implications of the nonconservation of the energy-momentum tensor by using the formalism of irreversible thermodynamics of open systems in the presence of matter creation/annihilation. The particle creation rates, pressure, and the expression of the comoving entropy are obtained in a covariant formulation and discussed in detail. Applied together with the gravitational field equations, the thermodynamics of open systems lead to a generalization of the standard $\Lambda$CDM cosmological paradigm, in which the particle creation rates and pressures are effectively considered as components of the cosmological fluid energy-momentum tensor. Furthermore, we also briefly present the coupling of curvature to geometry in conformal quadratic Weyl gravity, by assuming a coupling term of the form $L_m\tilde{R}^2$, where $L_m$ is the ordinary matter Lagrangian, and $\tilde{R}$ is the Weyl scalar. The coupling explicitly satisfies the requirement of the conformal invariance of the theory. Expressing $\tilde{R}^2$ with the use of an auxiliary scalar field and of the Weyl scalar, the gravitational action can be linearized in the Ricci scalar, leading in the Riemann space to a conformally invariant $f\left(R,L_m\right)$ type theory, with the matter Lagrangian nonminimally coupled to geometry.
\end{abstract}


\bodymatter

\section{Introduction}\label{aba:sec1}

The perplexing fact of the late-time cosmic acceleration \cite{SupernovaCosmologyProject:1998vns,SupernovaSearchTeam:1998fmf} has forced theorists and experimentalists to pose the question if General Relativity (GR) is the correct relativistic theory of gravitation. The fact GR is facing so many challenges, such as: (i) the difficulty in explaining particular observations; (ii) the incompatibility with other well established theories; (iii) and the lack of uniqueness, poses the question if this is indicative of a need for new gravitational physics? Cosmology is also an ideal testing ground for GR, in particular, the cause of the late-time cosmic acceleration, where a promising approach is to assume that at large scales GR breaks down, and a more general action describes the gravitational field. Indeed, many generalizations of the Einstein-Hilbert Lagrangian have been explored in the literature \cite{Nojiri:2010wj,Lobo:2008sg,Capozziello:2011et,Nojiri:2006ri,Lobo:2014ara,Avelino:2016lpj,CANTATA:2021ktz}, with the inclusion of several of the following invariants: $R^{2}$, $R_{\mu \nu }R^{\mu \nu }$, $R_{\alpha\beta \mu \nu }R^{\alpha \beta \mu \nu }$, $\varepsilon ^{\alpha\beta \mu \nu }R_{\alpha \beta \gamma \delta }R_{\mu \nu }^{\gamma\delta }$, $C_{\alpha \beta \mu \nu }C^{{\alpha \beta \mu \nu}}$, etc.

The physical motivations for these modifications of gravity include the possibility of a more
realistic representation of the gravitational fields near curvature singularities, and to create some first order approximation for the quantum theory of gravitational fields. Consider $f(R)$ gravity \cite{Sotiriou:2008rp}, for simplicity:
\begin{equation}
S=\int d^4x \,\sqrt{-g} \left[ \frac{f(R)}{2\kappa^2} + L_m(g^{\mu\nu}, \psi)  \right] \,,
\end{equation}
which combines an appealing feature, namely, it possesses mathematical simplicity and a fair amount of generality. Here, the Ricci scalar is a dynamical degree of freedom, which arises from the trace of the gravitational field equation, $3 \Box F + F R -2f = \kappa T$, where $F=df/dR$. This introduces a new light scalar degree of freedom and produces a late-time cosmic acceleration. But, the light scalar strongly violates the Solar System constraints. However, the way out to this problematic feature is through the `chameleon' mechanism, i.e., the scalar field becomes massive in the Solar System. Now, several approaches have been explored in the literature, namely, the metric \cite{Sotiriou:2008rp}, Palatini \cite{Olmo:2011uz}, metric-affine formalisms and the hybrid metric-Palatini formalism \cite{Harko:2011nh,Capozziello:2015lza,Harko:2018ayt}. Another interesting approach to modified gravity consists in an explicit nonminimal coupling between geometry and matter \cite{Bertolami:2007gv,Harko:2010mv,Harko:2014gwa,Harko:2020ibn,Harko:2011kv,Haghani:2013oma,Odintsov:2013iba}. In fact, this nonminimal coupling induces a non-conservation of the energy-momentum tensor, that implies non-geodesic motion and consequently leads to the appearance of an extra force. 

Here, we explore the physical and cosmological implications of the non-conservation of the energy-momentum tensor by using the formalism of irreversible thermodynamics of open systems in the presence of matter creation/annihilation, Indeed, we show that curvature-matter couplings offer a natural framework for irreversible thermodynamics of open systems and gravitationally induced matter creation processes. The systematic investigation of irreversible matter creation processes in General Relativity and cosmology started with the pioneering work by Prigogine and collaborators \cite{Pri0,Pri}. The description of particle creation is based on the reinterpretation of the energy-momentum tensor in Einstein's equations, by modifying the usual adiabatic energy conservation laws, and including an irreversible matter creation. Thus, matter creation corresponds to an irreversible energy flow from the gravitational field to the created matter constituents.

Even though at a phenomenological and formal level curvature-matter couplings may be introduced in a search for the maximal extension of the Hilbert-Einstein action, with the help of the substitution $\left(1/\kappa ^2\right)R+L_m\rightarrow f\left(R,L_m\right)$ \cite{Harko:2010mv}, where $f$ is an arbitrary analytical function of its arguments, finding a {\it physical principle} that would justify such a drastic change in the gravitational action would be of fundamental importance. Such a principle does indeed exist, and it is the requirement of the conformal invariance of the physical laws. The idea of the conformal invariance of natural phenomena was first introduced in the works of Herman Weyl \cite{Weyl, Weyl1} (see \cite{Scholz} for a detailed historical presentation of the evolution of Weyl's ideas). This will also be briefly  presented in this work.

In this proceedings, we will explore the need for new gravitational physics, focus on going
beyond GR, and analyse a specific extensions of $f(R)$ gravity, namely, on curvature-matter couplings and apply this theory to gravitationally induced particle creation.

\section{Nonminimal curvature-matter coupling}

\subsection{Action and field equations}

In this work, we explore a generalization of $f(R)$ gravity that includes a nonminimal curvature-matter coupling \cite{Bertolami:2007gv}
\begin{equation}
S=\int \left\{\frac{1}{2}f_1(R)+\left[1+\lambda f_2(R)\right]{
L}_{m}\right\} \sqrt{-g}\;d^{4}x~,
\label{action2}
\end{equation}
where $f_i(R)$ (with $i=1,2$) are arbitrary functions of the Ricci scalar $R$, and ${L}_{m}$ is the matter Lagrangian density. Many applications in astrophysics and cosmology have been explored in the literature (we refer the reader to \cite{Harko:2018ayt} for more details).

Varying the action (\ref{action2}) with respect to the metric $g_{\mu \nu }$ yields:
\begin{eqnarray}
F_1(R)R_{\mu \nu }-\frac{1}{2}f_1(R)g_{\mu \nu }-\nabla_\mu
\nabla_\nu \,F_1(R)+g_{\mu\nu}\square F_1(R) =-2\lambda F_2(R){L}_m R_{\mu\nu}
\nonumber \\
+2\lambda(\nabla_\mu
\nabla_\nu-g_{\mu\nu}\square){L}_m F_2(R)
+[1+\lambda f_2(R)]T_{\mu \nu }^{(m)}~,
\label{field}
\end{eqnarray}
where $F_i(R)=f'_i(R)$, with $'=d/dR$.

The matter energy-momentum tensor is defined as
\begin{equation}
T_{\mu \nu
}^{(m)}=-\frac{2}{\sqrt{-g}}\frac{\delta(\sqrt{-g}\,{L}_m)}{\delta(g^{\mu\nu})} ~.
\end{equation}

\subsection{Equation of motion for a perfect fluid}

Now, taking into account the covariant derivative of the field equation, the Bianchi identities, $\nabla^\mu
G_{\mu\nu}=0$, and the identity
\begin{equation}
(\square\nabla_\nu -\nabla_\nu\square)F_i=R_{\mu\nu}\,\nabla^\mu F_i ~,
\end{equation}
one finally deduces the relationship
\begin{equation}
\nabla^\mu T_{\mu \nu }^{(m)}=\frac{\lambda F_2}{1+\lambda
f_2}\left[g_{\mu\nu}{L}_m- T_{\mu \nu
}^{(m)}\right]\nabla^\mu R ~. \label{cons1}
\end{equation}
The nonminimal coupling between the matter and the higher derivative curvature terms implies the non-conservation of the energy-momentum tensor, which describes an exchange of energy and momentum between both. Note that analogous couplings arise after a conformal transformation in scalar-tensor theories of gravity (and string theory). In the absence of the coupling, one verifies the conservation of the energy-momentum, which is also induced from the diffeomorphism invariance of the matter part of the action.

In order to test the motion in the model, consider a perfect fluid. The equation of motion for a fluid element takes the following form
\begin{equation}
\frac{Du^{\alpha }}{ds} \equiv \frac{du^{\alpha }}{ds}+\Gamma _{\mu
\nu }^{\alpha }u^{\mu }u^{\nu }=f^{\alpha }~, \label{eq1}
\end{equation}
where $f^\alpha$ is an extra-force, given by
\begin{eqnarray}
\label{force}
f^{\alpha }&=&\frac{1}{\rho +p}\Bigg[\frac{\lambda
F_2}{1+\lambda f_2}\left({L}_m-p\right)\nabla_\nu
R+\nabla_\nu p \Bigg] h^{\alpha \nu }\,.
\end{eqnarray}
$h_{\mu \lambda }=g_{\mu \lambda }+u_{\mu }u_{\lambda }$ is the projection operator;
$f^{\alpha }$ is orthogonal to the 4-velocity of the particle, $f^{\alpha }u_{\alpha }=0$.
An intriguing feature is that depending on the choices of the thermodynamic matter Lagrangian given by ${L}_m=p$ or ${L}_m=-\rho$, which are equivalent in GR, yield different dynamics in the presence of a nonminimal curvature-matter coupling \cite{Bertolami:2008ab}.

\subsection{Scalar-tensor representation}

The analysis is rather simplified in the scalar-tensor formulation of the theory, where one introduces
two new fields $\phi$ and $\psi(\phi) \equiv f_1^{\prime}(\phi) $, and the action~(\ref{action2}) can be reformulates as
\begin{equation}  \label{300}
S=\int d^4x \sqrt{-g} \left[ \frac{\psi R }{2} -V(\psi)\, +U(\psi) L_m %
\right] ,
\end{equation}
where the two potentials $V(\psi)$ and $U(\psi)$ of the theory are given by:
\begin{equation}  \label{400}
V(\psi) = \frac{\phi(\psi) f_1^{\prime }\left[ \phi (\psi ) \right] -f_1%
\left[ \phi( \psi ) \right] }{2},
\end{equation}
\begin{equation}  \label{500}
U( \psi) = 1+\lambda f_2\left[ \phi( \psi ) \right] ,
\end{equation}
respectively. The function $\phi (\psi)$ is obtained by inverting $%
\psi(\phi) \equiv f_1^{\prime }(\phi) $.  The actions~(\ref{action2})
and (\ref{300}) are equivalent when $f_1^{\prime \prime }(R) \neq 0$.

In the scalar-tensor representation, the divergence of the energy-momentum tensor is given by
\begin{equation}
\nabla _{\mu }T_{\nu }^{\mu }=-\left[ \nabla _{\mu }\ln U\left( \psi \right) %
\right] T_{\nu }^{\mu }-\frac{2V^{\prime }\left( \psi \right) -U^{\prime
}(\psi )L_{m}}{U(\psi )}\nabla _{\nu }\psi .  \label{17}
\end{equation}
Assuming a perfect fluid, the energy balance equation is:
\begin{equation}  \label{eneq}
\dot{\rho}+3H(\rho +p)+\rho \frac{d}{ds}\ln U(\psi )+\frac{2V^{\prime
}\left( \psi \right) -U^{\prime }(\psi )L_{m}}{U(\psi )}\dot{\psi}=0.
\end{equation}%

\section{Thermodynamics of open systems}

\subsection{Irreversible matter creation processes}

The systematic investigation of irreversible matter creation processes in General Relativity and cosmology started with the pioneering work by Prigogine and collaborators
\cite{Pri0,Pri}. The description of particle creation is based on the reinterpretation of the energy-momentum tensor in Einstein's equations, by modifying the usual adiabatic energy conservation laws, and including an irreversible matter creation. Thus, matter creation corresponds to an irreversible energy flow from the gravitational field to the created matter constituents.

Consider the flat isotropic and homogeneous FLRW metric:
\begin{equation}
ds^2=dt^2-a^2(t)\left(dx^2+dy^2+dz^2\right).
\end{equation}
Taking into account the thermodynamical implications, where one considers that the Universe contains $N$ particles in a volume $V$, an energy density $\rho $ and a thermodynamic pressure $p$. For this cosmological system, the 1st law of thermodynamics, in its most general form, is given by \cite{Pri0}
\begin{equation}
\frac{d}{dt}\left( \rho a^{3}\right) +p\frac{d}{dt}a^{3}=\frac{dQ}{dt}+\frac{%
\rho +p}{n}\frac{d}{dt}\left( na^{3}\right) ,  \label{21}
\end{equation}%
where $dQ$ is the heat received by the system during time $dt$, and $n=N/V$ is the particle number density.
	
We restrict our analysis to adiabatic transformations defined by the condition $dQ=0$, that is, ignore proper heat transfer processes in the cosmological system. Thus:
\begin{equation}
\dot{\rho}+3(\rho +p)H=\frac{\rho +p}{n}\left( \dot{n}+3Hn\right) .
\label{cons0}
\end{equation}
Hence, in the irreversible thermodynamics of open systems, one can consider that the ``heat''
(internal energy), received/lost by the system is due to the change in the particle number $n$.

The time variation of the particle number density obtained as
\begin{equation}
\dot{n}+3nH=\Gamma n, \label{22}
\end{equation}%
where $\Gamma$ is the particle creation rate.
Therefore, the energy conservation equation can be reformulated in the
alternative form
\begin{equation}
\dot{\rho}+3(\rho +p)H=(\rho +p)\Gamma .  \label{41}
\end{equation}

For adiabatic transformations describing irreversible particle creation in an open thermodynamic systems, the 1st law of thermodynamics can be rewritten as an effective energy conservation equation,
\begin{equation}
\frac{d}{dt}\left( \rho a^{3}\right) +\left( p+p_{c}\right) \frac{d}{dt}%
a^{3}=0,
\end{equation}%
or, in an equivalent form, as,
\begin{equation}
\dot{\rho}+3\left( \rho +p+p_{c}\right) H=0,  \label{comp}
\end{equation}%
where we have introduced a new thermodynamic quantity, $p_{c}$, denoted the
creation pressure and defined as
\begin{eqnarray}  \label{pc1}
p_{c} = -\frac{\rho +p}{3}\frac{\Gamma }{H}.
\end{eqnarray}

The inclusion of the matter creation processes into the Einstein equations lead to the possibility of cosmological models that start from empty conditions and gradually build up matter and entropy. Note that gravitational entropy takes a simple meaning as associated with the entropy that is necessary to produce matter. The matter creation is described macroscopically by introducing a negative pressure due to matter creation. Thus, cosmological particle creation can take place from the quantum vacuum, due to external conditions, which are caused by the expansion or contraction of the Universe \cite{Q1,Q3,Q5}.

We will explore the possibility that modified theories of gravity with a curvature-matter coupling can provide a phenomenological description of particle production in the cosmological fluid filling the Universe \cite{Harko:2015pma}.

\section{Gravitationally induced particle creation}

We consider the physical interpretation of the curvature-matter coupling by adopting the point of view of the thermodynamics of open systems, in which matter creation irreversible processes may take place at a cosmological scale. The energy conservation equation contains, as compared to the standard adiabatic conservation equation, an extra term, which can be interpreted in the framework of the open thermodynamic systems as an irreversible matter creation rate.
According to irreversible thermodynamics, matter creation also represents an
entropy source, generating an entropy flux, and thus leading, in the
presence of the curvature-matter coupling, to a modification in the
temperature evolution.

\subsection{Particle creation rate and creation pressure}

Therefore, from the point of view of the thermodynamics of open systems, the energy balance equation, in the presence of a curvature-matter coupling, can be interpreted as describing particle
creation in an homogeneous and isotropic geometry, with the time variation of the particle number density obtained as
\begin{equation}
\dot{n}+3nH=\Gamma n,  \label{22b}
\end{equation}%
where the particle creation rate $\Gamma $ is defined as
\begin{equation}  \label{33}
\Gamma =-\frac{1}{\rho +p}\Bigg\{\rho \frac{d}{dt}\ln U(\psi )+\frac{%
2V^{\prime }\left( \psi \right) -U^{\prime }(\psi )L_{m}}{U(\psi )}\dot{\psi}%
\Bigg\}.
\end{equation}

Therefore, the energy conservation equation can be reformulated in the
alternative form
\begin{equation}
\dot{\rho}+3(\rho +p)H=(\rho +p)\Gamma .  \label{41b}
\end{equation}

As proven for adiabatic transformations, the 1st law describing irreversible particle creation in an open thermodynamic systems, can be rewritten as an effective energy conservation equation,
\begin{equation}
\frac{d}{dt}\left( \rho a^{3}\right) +\left( p+p_{c}\right) \frac{d}{dt}%
a^{3}=0,
\end{equation}%
or, in an equivalent form, as,
\begin{equation}
\dot{\rho}+3\left( \rho +p+p_{c}\right) H=0,  \label{compb}
\end{equation}%
where the the creation pressure, $p_{c}$, is defined as
\begin{eqnarray}  \label{pc1b}
p_{c} = -\frac{\rho +p}{3}\frac{\Gamma }{H}.
\end{eqnarray}%

Therefore, the creation pressure, with a curvature-matter coupling, is given by
\begin{equation}  \label{pc}
p_{c}=-\frac{1}{3H}\left[\rho \frac{d}{dt}\ln U(\psi )+\frac{2V^{\prime
}\left( \psi \right) -U^{\prime }(\psi )L_{m}}{U(\psi )}\dot{\psi}\right].
\end{equation}

\subsection{Entropy and temperature evolution}

The basic principles of the thermodynamics of open systems state that the
entropy change consists of two components: $d_eS=$entropy flow term, $d_iS=$entropy creation term.  The total entropy $S$ of an open thermodynamic system can represented as \cite{Pri0,Pri}:
\begin{equation}
dS = d_eS + d_iS,
\end{equation}
where by definition $d_iS > 0$. In the case of a closed thermodynamic system and for adiabatic
transformations, we have $dS=0$ and $d_iS=0$. However, with a curvature-matter coupling, leading to effective matter creation, there is a non-zero contribution to $S$.

For a homogeneous and isotropic Universe the entropy flow term, we have $d_eS = 0$. But, matter creation represents a source for entropy creation,
and the time variation $d_i S$ is \cite{Pri}:
\begin{eqnarray}  \label{25}
T\frac{d_iS}{dt}=T\frac{dS}{dt}=\frac{h}{n}\frac{d}{dt}\left(na^3\right)-%
\mu \frac{d}{dt}\left(na^3\right)    \geq 0.
\end{eqnarray}
Equation~(\ref{25}) gives the time variation of the entropy as
\begin{equation}
\frac{dS}{dt}=\frac{S}{n}\left( \dot{n}+3Hn\right) =\Gamma S\geq 0,
\label{43}
\end{equation}%
so that the entropy increase due to particle production yields
\begin{equation}
S(t)=S_{0}e^{\int_{0}^{t}{\Gamma \left( t^{\prime }\right) dt^{\prime }}},
\end{equation}
where $S_{0}=S(0)$ is a constant.

With the use of Eq.~(\ref{43}), we obtain
for the entropy creation in the scalar-tensor representation of the linear
coupling between matter and geometry the following equation
\begin{equation}
\frac{1}{S}\frac{dS}{dt}=-\frac{1}{\rho +p}\Bigg\{\rho \frac{d}{dt}\ln
U(\psi )+\frac{2V^{\prime }\left( \psi \right) -U^{\prime }(\psi )L_{m}}{%
U(\psi )}\dot{\psi}\Bigg\}.  \label{entff}
\end{equation}

The entropy flux four-vector $S^{\mu }$ is defined as
\begin{equation}
S^{\mu }=n\sigma U^{\mu },
\end{equation}%
where $\sigma =S/N$ is the specific entropy per particle.
$S^{\mu }$ must
satisfy the second law of thermodynamics, which imposes the constraint $%
\nabla _{\mu }S^{\mu }\geq 0$.
Using the Gibbs relation and the definition of the chemical potential $\mu$, yields
\begin{eqnarray}
\nabla _{\mu }S^{\mu } = \frac{1}{T}\left( \dot{n}+3Hn\right) \left( \frac{h}{n}-\mu \right) ,
\end{eqnarray}%
The entropy production rate due to the particle creation
processes, with the nonminimal curvature-matter coupling, is given by
\begin{eqnarray}
\nabla _{\mu }S^{\mu }=
\frac{n}{T(\rho +p)}\Bigg\{\rho \frac{d}{dt}\ln U(\psi )
+\frac{2V^{\prime }\left( \psi \right) -U^{\prime }(\psi )L_{m}}{U(\psi )}%
\dot{\psi}\Bigg\}\left( \frac{h}{n}-\mu \right) . \nonumber
\end{eqnarray}

A general thermodynamic system is described by two fundamental thermodynamic
variables, the particle number density $n$, and the temperatures $T$,
respectively. If the system is in an equilibrium state, the energy density $%
\rho $ and the thermodynamic pressure $p$ are obtained, in terms of $n$ and $%
T$, from the equilibrium equations of state of the matter,
\begin{equation}
\rho =\rho (n,T), \qquad p=p(n,T).  \label{51}
\end{equation}%
Therefore the energy conservation equation can be obtained in
the general form
\begin{equation}
\frac{\partial \rho }{\partial n}\dot{n}+\frac{\partial \rho }{\partial T}%
\dot{T}+3(\rho +p)H=\Gamma n.
\end{equation}%

Using the general thermodynamic relation
\begin{equation}
\frac{\partial \rho }{\partial n}=\frac{h}{n}-\frac{T}{n}\frac{\partial p}{%
\partial T},
\end{equation}%
the temperature evolution of the newly created particles due
to the curvature-matter coupling is given by the expression
\begin{equation}
\frac{\dot{T}}{T}=c_{s}^{2}\frac{\dot{n}}{n}=c_{s}^{2}\left( \Gamma
-3H\right) ,  \label{54}
\end{equation}%
where the speed of sound $c_{s}$ is defined as $c_{s}^{2}=\partial
p/\partial \rho $.

If the geometrically created matter satifies a barotropic
equation of state of the form $p=\left( \gamma -1\right) \rho $, $1\leq
\gamma \leq 2$, the temperature evolution follows the simple equation
\begin{equation}
T=T_{0}n^{\gamma -1}.
\end{equation}

\subsection{Validity of the second law of thermodynamics}

Here, we consider the validity of the second law of thermodynamics in
cosmology. If one defines the entropy, $S$, measured by a
comoving observer as the entropy of the apparent horizon plus that
of matter and radiation inside it, then the Universe approaches
thermodynamic equilibrium as it nears the final de Sitter phase.
Then it follows that $S$ increases, and that it is concave, thus
leading to the result that the second law of thermodynamics is
valid, as one should expect given the strong connection between
gravity and thermodynamics, for the case of the expanding
Universe.

For a spatially-flat FRW Universe filled with dust, one has
\begin{equation}
S = S_{ah} + S_m = \frac{\pi}{H^2} + \frac{4\pi}{3H^3}\,n(t)\; ,
\label{MP}
\end{equation}
where we have used the fact that the radius of the apparent horizon is $%
r_{ah}= H^{-1}$.
Therefore, the thermodynamic requirements
$S^{\prime}\geq 0 $ and $S^{\prime \prime }\leq 0$ impose specific
constraints on the particle creation rate $\Gamma $, and its
derivative with respect to the scale factor.

A particularly important case is that of the de Sitter evolution of the Universe,
with $H=H_{\star}={\rm constant}$. In this case, we have
\begin{equation}
S^{\prime }= \frac{4\pi }{ 3H_{\star}^{4}}\frac{\left[ \Gamma (a)
-3H_{\star} \right]}{a}n(a) \geq 0  \,,
\end{equation}
and
\begin{eqnarray}
S^{\prime \prime}= \frac{4 \pi n(a)}{3 a^2 H_{\star}^5}
\Big\{\Gamma ^2(a)+H_{\star} \left[a \Gamma ^{\prime }(a)+12
H_{\star}\right] -7 H_{\star} \Gamma (a)\Big\} \leq 0 \,,
\end{eqnarray}
respectively.

Accordingly, the constraints
\begin{equation}
\Gamma \geq 3H_{\star}, \qquad  \Gamma ^{\prime
}(a)\leq \left[7\Gamma (a)-\Gamma
^2(a)/H_{\star}-12H_{\star}\right]/a \,,
\end{equation}
are imposed on the particle creation
rate $\Gamma$.
For $\Gamma =3H_{\star}$, we obtain $S =
const$, showing that in this case the cosmological
evolution is isentropic.
Here, $H_{\star}$ denotes the expansion
rate of the final de Sitter phase.

\subsection{Cosmological applications}

 Using the scalar-tensor representation of the theory, we have obtained the particle creation rate, the creation pressure and the entropy associated to the gravitational energy transfer to matter.
The gravitational field equations corresponding to these choices have a de Sitter type accelerating solution.
This cosmic acceleration is triggered by the particle creation process, which generates a negative creation pressure.
The late-time cosmic acceleration may be considered as an empirical evidence for matter creation, and a viable alternative to the mysterious dark energy.

The potentials $V(\psi)$ and $U(\psi)$, that characterize the curvature-matter coupling, can be provided by fundamental quantum field theoretical models of the gravitational interaction:
this opens the possibility of an in-depth comparison of the predictions of the theory with cosmological and astrophysical observational data \cite{Harko:2015pma}.

\subsection{Extended geometry-matter couplings}

\subsubsection{$f(R,L_m)$ gravity}

The linear curvature-matter coupling theory was generalized in the so-called $f(R,L_m)$ gravity \cite{Harko:2010mv}. The gravitational field equations in the metric formalism, as well as the equations of motion for test particles, which follow from the covariant divergence of the energy-momentum tensor, were obtained. Generally, the motion is non-geodesic, and takes place in the presence of an extra force orthogonal to the four-velocity, as in the linear curvature-matter coupling analysed above. The gravitational field equations and the equations of motion for a particular model in which the action of the gravitational field has an exponential dependence on the standard general relativistic Hilbert--Einstein Lagrange density were also derived.

The action is given by the following action \cite{Harko:2010mv}:
\begin{equation}
S=\int f\left(R,L_m\right) \sqrt{-g}\;d^{4}x\,.
   \label{action}
\end{equation}
The gravitational field equation is given by:
\begin{eqnarray}\label{fieldb}
&&f_{R}\left( R,L_{m}\right) R_{\mu \nu }+\left( g_{\mu \nu
}\nabla _{\mu }\nabla^{\mu } -\nabla
_{\mu }\nabla _{\nu }\right) f_{R}\left( R,L_{m}\right)
	\nonumber\\
&&-\frac{1}{2}\left[ f\left( R,L_{m}\right) -f_{L_{m}}\left(
R,L_{m}\right)L_{m}\right] g_{\mu \nu }=\frac{1}{2}
f_{L_{m}}\left( R,L_{m}\right) T_{\mu \nu }\,.
\end{eqnarray}
Note that for the Hilbert-Einstein Lagrangian: $f( R,L_{m})=R/2\kappa^2+L_{m}$, we recover the standard Einstein field equations. For $f\left( R,L_{m}\right)=f_1(R)+\left[1+\lambda f_2(R)\right]{
L}_{m} $, we recover the field equations with an arbitrary linear curvature-matter coupling. We refer the reader to \cite{Harko:2010mv,Harko:2014gwa,Harko:2020ibn} for more details.

\subsubsection{$f(R,T)$ gravity}

Another related theory is $f(R,T)$ gravity, which is given by the following action \cite{Harko:2011kv}
\begin{equation}
S=\frac{1}{16\pi}\int
f\left(R,T\right)\sqrt{-g}\;d^{4}x+\int{L_{m}\sqrt{-g}\;d^{4}x}\,.
\end{equation}
where $f\left(R,T\right)$ is an arbitrary function of the Ricci scalar,
$R$, and of the trace $T$ of the energy-momentum tensor of the
matter, $T_{\mu \nu}$.

Note that the dependence from $T$ may be induced by exotic imperfect fluids or quantum effects (conformal anomaly). This theory may also be considered a relativistically covariant model of interacting dark energy. We refer the reader to \cite{Harko:2011kv,Harko:2018ayt} for further details.

\subsubsection{Further generalization: $f(R,T,R_{\mu\nu}T^{\mu\nu})$ gravity}

A further generalization to $f(R,T)$ gravity was explored in \cite{Haghani:2013oma,Odintsov:2013iba} in the so-called $f(R,T,R_{\mu\nu}T^{\mu\nu})$ gravity, where the effective Lagrangian of the gravitational field is given by an arbitrary function of the Ricci scalar, $R$ the trace of the matter energy-momentum tensor, $T$, and the contraction of the Ricci tensor $R_{\mu\nu}$ with the matter energy-momentum tensor, $T_{\mu\nu}$.

In \cite{Haghani:2013oma}, the theory was presented in detail, and the Newtonian limit of the model was also considered, and an explicit expression for the extra-acceleration which depends on the matter density was obtained in the small velocity limit for dust particles. The Dolgov-Kawasaki instability was analysed, and the stability conditions of the model with respect to local perturbations was obtained. A particular class of gravitational field equations was also be obtained by imposing the conservation of the energy-momentum tensor, where the corresponding field equations was derived  for the conservative case by using a Lagrange multiplier method, from a gravitational action that explicitly contains an independent parameter multiplying the divergence of the energy-momentum tensor. The cosmological implications of the model were investigated for both the conservative and non-conservative cases, and several classes of analytical solutions are obtained. We refer the reader to \cite{Haghani:2013oma,Odintsov:2013iba} for more details.

\section{Future outlook: conformal invariance, Weyl geometry and curvature-matter coupling}

The idea of the conformal invariance of natural phenomena was first introduced in the works of Herman Weyl \cite{Weyl, Weyl1} (see \cite{Scholz} for a detailed historical presentation of the evolution of Weyl's ideas). Weyl's approach to gravitational phenomena is based on the observation that Maxwell's equations in vacuum are conformally invariant. Hence, if the laws of nature are unitary, it is natural to suggest that the gravitational field equations as developed by Einstein must have the same symmetry. In order to systematically implement the idea of conformal invariance Weyl constructed a  geometry in which the covariant derivative of the metric tensor does not identically vanish, and has the property $\nabla _{\lambda}g_{\mu \nu}=Q_{\lambda \mu \nu}=\omega_{\lambda}g_{\mu \nu}$, where $Q_{\lambda \mu \nu}$ is the nonmetricity, while $\omega _{\lambda}$ is the Weyl vector field \cite{Weyl1}.  In Weyl geometry the parallel transport of a vector does not maintain its length constant. Due to this geometric property Einstein criticized the physical interpretation of the new geometry as proposed by Weyl, pointing out that if the running of the atomic clocks would depend on their past evolution, it would be impossible for the sharp spectral lines to exist in the presence of an electromagnetic fields.

However, even though Weyl's geometry has been abandoned as a unified candidate for a unified field theory, the idea of the conformal invariance of the physical laws proved to be very attractive, and it opened some new perspectives on the interpretation and foundations of some physical theories. In particular, {\it if one requires the conformal invariance of the gravitational action in the presence of matter}, {\it the only possibility to add a conformally invariant matter term would be through a curvature matter coupling}. In the following we will briefly present a particular gravitational model with geometry-matter coupling, in which conformal invariance uniquely fixes the form of the coupling term.

\subsection{Weyl geometry in a nutshell}

We define Weyl geometry as the classes of equivalence $\left( g_{\alpha
\beta },\omega _{\mu }\right) $ of the metric $g_{\alpha \beta }$ and of the
Weyl vector gauge field $\omega _{\mu }$. The two elements of the equivalence class are related by the Weyl gauge
transformations \cite{Gh7},
\begin{equation}
\tilde{g}_{\mu \nu } =\Sigma ^{n}g_{\mu \nu }=\left[\tilde{g}_{\mu \nu}\right],\quad \tilde{\omega}_{\mu }=\omega
_{\mu }-\frac{1}{\alpha }\partial _{\mu }\ln \Sigma , \quad 
\sqrt{-\tilde{g}} =\Sigma ^{2n}\sqrt{-g},\tilde{\phi}=\Sigma ^{-n/2}\phi ,
\end{equation}
where $n$ is the Weyl charge. We also have
\begin{eqnarray*}
\left[ \tilde{R}_{\mu \nu }\right] &=&1,\left[ \tilde{\Gamma}_{\nu \rho }^{\mu
}\right] =1,\left[ \tilde{R}\right] =\frac{1}{\Sigma ^n},\left[ \tilde{R}_{\nu \rho \sigma
}^{\mu }\right] =1,\left[ F_{\mu \nu }\right] =1,
\left[ L_{m}\right] =1,\left[ T_{\mu \nu }\right] =\Sigma^n ,\nonumber\\
\left[ T^{\mu \nu
}\right] &=&\Sigma ^{-n},\left[ \rho \right] =1,\left[ p\right] =1,
\left[ T%
\right] =1, \left[ u_{\mu }\right] =\Sigma ^{n/2},\left[ u^{\mu }\right] =\Sigma ^{-n/2},%
\left[ j^{\mu }\right] =\Sigma ^{-n/2},
\end{eqnarray*}
where the square brackets $[...]$
denote the degree of $\Sigma$ in the conformal transformation of the geometrical and physical quantities.

The Weyl gauge vector field can be obtained from the Weyl connection $\tilde{\Gamma}
$, which is defined as a solution of the system of equations
\begin{equation}\label{5}
\tilde{\nabla}_{\lambda }g_{\mu \nu }=-n\alpha \omega _{\mu }g_{\mu \nu },
\end{equation}%
where $\alpha$ is the Weyl gauge coupling. By using the standard definition of the covariant derivative we find
\begin{equation}
\tilde{\nabla}_{\lambda }g_{\mu \nu }=\partial _{\lambda }g_{\mu \nu }-%
\tilde{\Gamma}_{\nu \lambda }^{\rho }g_{\rho \mu }-\tilde{\Gamma}_{\mu
\lambda }^{\rho }g_{\nu \rho }.
\end{equation}
One of the important properties of Weyl geometry is its {\it non-metric} nature. Equivalently, Eq.~(\ref{5}) can be reformulated as
$
\left( \tilde{\nabla}_{\lambda }+n\alpha \omega _{\lambda }\right) g_{\mu
\nu }=0.
$

We can obtain gauge invariant expressions in Weyl geometry by replacing the ordinary derivative with Weyl covariant derivative. For example, the expression obtained through the substitution
$
\partial _{\lambda }\rightarrow \partial _{\lambda }+\mathrm{weight}\times
\alpha \times \omega _{\lambda }
$
is gauge invariant.
From Eq.~(\ref{5}) we obtain the expression of the Weyl connection as
\begin{equation}
\tilde{\Gamma}_{\mu \nu }^{\lambda }=\Gamma _{\mu \nu }^{\lambda }+\alpha
\frac{n}{2}\left( \delta _{\mu }^{\lambda }\omega _{\nu }+\delta _{\nu
}^{\lambda }\omega _{\mu }-\omega ^{\lambda }g_{\mu \nu }\right) ,
\label{1a}
\end{equation}%
where
\begin{equation}
\Gamma _{\lambda ,\mu \nu }=\frac{1}{2}\left( \partial _{\nu }g_{\lambda \mu
}+\partial _{\mu }g_{\lambda \nu }-\partial _{\lambda }g_{\mu \nu }\right) ,
\end{equation}%
is the Levi-Civita metric connection, and
$
\tilde{\Gamma}_{\mu \nu }^{\lambda }=g^{\lambda \sigma }\tilde{\Gamma}%
_{\lambda ,\mu \nu }.
$
The trace of Eq. (\ref{1a}) is given by
$
\tilde{\Gamma}_{\mu }=\Gamma _{\mu }+2n\alpha \omega _{\mu }.
$

The field strength $\tilde{F}_{\mu \nu }$ associated to the Weyl vector $\omega _{\mu
}$ is defined as
\begin{equation}
\tilde{F}_{\mu \nu }=\tilde{\nabla}_{\mu }\omega _{\nu }-\tilde{\nabla}_{\nu
}\omega _{\mu }=\partial _{\mu }\omega _{\nu }-\partial _{\nu }\omega _{\mu
}.
\end{equation}

With the help of the Weyl connection we can construct the curvature tensor
as follows,
\begin{equation}
\tilde{R}_{\mu \nu \sigma }^{\lambda }=\partial _{\nu }\tilde{\Gamma}_{\mu
\sigma }^{\lambda }-\partial _{\sigma }\tilde{\Gamma}_{\mu \nu }^{\lambda }+%
\tilde{\Gamma}_{\rho \nu }^{\lambda }\tilde{\Gamma}_{\mu \sigma }^{\rho }-%
\tilde{\Gamma}_{\rho \sigma }^{\lambda }\tilde{\Gamma}_{\mu \nu }^{\rho },
\end{equation}%
and its first contraction,
\begin{equation}
\tilde{R}_{\mu \nu }=\tilde{R}_{\mu \lambda \nu }^{\lambda },\tilde{R}%
=g^{\mu \sigma }\tilde{R}_{\mu \sigma },
\end{equation}%
respectively. The Weyl scalar is given by
\begin{equation}
\tilde{R}=R-3n\alpha \nabla _{\mu }\omega ^{\mu }-\frac{3}{2}\left( n\alpha
\right) ^{2}\omega _{\mu }\omega ^{\mu }.  \label{R}
\end{equation}

It is easy to check that $\tilde{R}$ transforms covariantly, and $\sqrt{-g}\tilde{R}^{2}$ is
invariant with respect to the gauge transformations. The Weyl tensor, an important geometric and physical quantity, can be obtained as
\begin{equation}
\tilde{C}_{\mu \nu \rho \sigma }^{2}=C_{\mu \nu \rho \sigma }^{2}+\frac{3}{2}%
\left( \alpha n\right) ^{2}\tilde{F}_{\mu \nu }^{2},
\end{equation}
where $C_{\mu \nu \rho \sigma }$ is the Weyl tensor defined in the Riemannian geometry \cite{LaLi}.
The tensor $\sqrt{-g}\tilde{C}_{\mu \nu \rho \sigma }^{2}$ is invariant with respect to the conformal transformations of the metric.

For $C_{\mu \nu \rho \sigma }^{2}$ we find
\begin{equation}
C_{\mu \nu \rho \sigma }^{2}=R_{\mu \nu \rho \sigma }R^{\mu \nu \rho \sigma
}-2R_{\mu \nu }R^{\mu \nu }+\frac{1}{3}R^{2}.
\end{equation}

For the Weyl charge $n$ in the following we will adopt the value $n=1$.

\subsection{Coupling curvature and matter in Weyl geometry}

With the help of the fundamental scalars of Weyl geometry $\left( \tilde{R},\tilde{F}%
_{\mu \nu }^{2},\tilde{C}_{\mu \nu \rho \sigma }^{2}\right) $, one can construct the 
conformally invariant gravitational action \cite{Gh7,Gh4,Gh5,Gh6} 
gravitational field,
\begin{equation}  \label{S0}
S_{0}=\int \Big[\frac{1}{4!}\frac{1}{\xi ^{2}}\tilde{R}^{2}-\frac{1}{4}\,%
\tilde{F}_{\mu \nu }^{2}-\frac{1}{\eta ^{2}}\tilde{C}_{\mu \nu \rho \sigma
}^{2}\Big]\sqrt{-g}d^{4}x,
\end{equation}%
where $\xi $ and $\eta $ are dimensionless coupling parameters. However, to obtain physically realistic gravitational models the effect of
the matter must also be included in the action (\ref{S0}) by using a
conformally invariant Lagrangian density $\tilde{L}_{m}$.

The simplest possibility for adding a
{\it conformally invariant matter Lagrangian} is by adopting for the matter contribution the form $\tilde{L}%
_{m}= L_{m}\tilde{R}^{2}/4!\gamma ^2$, where $L_{m}$ is the ordinary matter
Lagrangian density, satisfying $\left[L_m\right]=1$, and $\gamma $ is a coupling constant. Hence we obtain for the conformally invariant action for gravity in Weyl
geometry the expression
\begin{eqnarray}\label{S1}
S&=&\int \Big[\frac{1}{4!\xi ^2}\tilde{R}^{2}-\frac{1}{4}\,\tilde{F}_{\mu
\nu }^{2}-\frac{1}{\eta ^{2}}\tilde{C}_{\mu \nu \rho \sigma }^{2}+\frac{1}{%
4!\gamma ^{2}}L_{m}\tilde{R}^{2}\Big]\sqrt{-g}d^{4}x  \notag  \label{S} \\
&=&\int \Big[\frac{1}{4!\xi^2}\left( 1+\frac{\xi ^{2}}{\gamma ^{2}}L_m\right)
\tilde{R}^{2}-\frac{1}{4}\,\tilde{F}_{\mu \nu }^{2}-\frac{1}{\eta ^{2}}%
\tilde{C}_{\mu \nu \rho \sigma }^{2}\Big]\sqrt{-g}d^{4}x.
\end{eqnarray}

By introducing now an auxiliary scalar field $\phi _{0}$, defined according to $\tilde{R}^{2}\equiv -2\phi _{0}^{2}\tilde{R}-\phi _{0}^{4}$ \cite{Gh7},
 after substitution in the action (\ref{S1}),  we obtain
\begin{eqnarray}
\hspace{-0.5cm}S &=&-\int \Bigg\{\frac{1}{2\xi ^{2}}\left( 1+\frac{\xi ^{2}}{%
\gamma ^{2}}L_{m}\right) \Bigg[\frac{\phi _{0}^{2}}{6}R-\frac{\alpha }{2}%
\phi _{0}^{2}\nabla _{\mu }\omega ^{\mu }  \notag \\
\hspace{-0.5cm} &&-\frac{\alpha ^{2}}{4}\phi _{0}^{2}\omega _{\mu }\omega
^{\mu }+\frac{\phi _{0}^{4}}{12}\Bigg]+\frac{1}{4}\,\tilde{F}_{\mu \nu }^{2}+%
\frac{1}{\eta ^{2}}\tilde{C}_{\mu \nu \rho \sigma }^{2}\Bigg\}\sqrt{-g}%
d^{4}x.  
\end{eqnarray}

We perform now a conformal transformation of the metric with the help of the conformal
factor $\Sigma =\phi _{0}^{2}/\left\langle \phi _{0}^{2}\right\rangle $,
where $\left\langle \phi _{0}^{2}\right\rangle $ is the (constant) vacuum
expectation value of the field $\phi _{0}$ \cite{Gh7}. The determinant of the
metric tensor transforms a $\sqrt{-g}=\left( \left\langle
\phi _{0}^{2}\right\rangle ^{2}/\phi _{0}^{4}\right) \sqrt{-\hat{g}}$, while the Ricci scalar becomes $%
\tilde{R}=\left( \phi _{0}^{2}/\left\langle \phi _{0}^{2}\right\rangle
\right) \hat{R}$. Moreover, the Weyl vector transforms as $\omega _{\mu }=\hat{\omega}_{\mu }+\left( 2/\alpha
\right) \partial _{\mu }\phi_0/\phi_0 $, and we also {\it impose the
gauge condition} $\nabla_{\mu }\hat{\omega}^{\mu }=0$.

 Hence, after using the gauge freedom of the theory, the Weyl geometry action containing a curvature-matter coupling becomes
\begin{align}\label{Sf1}
S &=-\int \Bigg\{\left( 1+\frac{\xi ^{2}}{\gamma ^{2}}L_{m}\right) \Bigg[%
\frac{1}{2}M_{p}^{2}R-\frac{3\alpha ^{2}}{4 }M_p^2\omega _{\mu }\omega ^{\mu
}  \notag \\
&+\frac{3}{2}\xi ^{2}M_{p}^{4}\Bigg] +\frac{1}{4\delta ^{2}}\,\tilde{F}%
_{\mu \nu }^{2}+\frac{1}{\eta ^{2}}C_{\mu \nu \rho \sigma }^{2}\Bigg\}\sqrt{%
	-g}d^{4}x,
\end{align}%
where $M_{p}^{2}=\left\langle \phi _{0}^{2}\right\rangle
/6\xi ^{2}$ and $1/\delta ^{2}=1+6\alpha ^{2}/\eta ^{2}$, and, to keep the notation simple, the hats are not written out explicitly on the conformally rescaled
geometrical quantities. It is important to point out that in action (\ref{Sf1}) the physical and
geometrical quantities are defined in the Riemann space.
We can further rescale the matter Lagrangian by defining a new effective matter variable ${L}_m$, given by
\begin{equation}
{L}_m=1+\frac{\xi ^2}{\gamma ^2}L_m.
\end{equation}
Thus, we obtain the action of the {\it conformally invariant} $f\left(R,L_m\right)$ theory as
\begin{align}\label{Sf}
S &=-\int \Bigg\{{L}_m \Bigg[%
\frac{1}{2}M_{p}^{2}R-\frac{3\alpha ^{2}}{4 }M_p^2\omega _{\mu }\omega ^{\mu
}  
+\frac{3}{2}\xi ^{2}M_{p}^{4}\Bigg] +\frac{1}{4\delta ^{2}}\,\tilde{F}%
_{\mu \nu }^{2}+\frac{1}{\eta ^{2}}C_{\mu \nu \rho \sigma }^{2}\Bigg\}\sqrt{%
	-g}d^{4}x.
\end{align}%

\subsection{Gravitational and Weyl field equations}

We vary first the gravitational action (\ref{Sf}) with respect to the Weyl vector $%
\omega_{\mu}$, and thus we obtain the generalized system of
Maxwell-Proca type equations for $\omega _{\mu}$, given by
\begin{equation}
\nabla _{\mu }\tilde{F}^{\mu \nu }+\frac{3}{%
	2}M_p^2\alpha ^{2}\delta^2\left( 1+\frac{\xi ^{2}}{\gamma ^{2}}L_{m}\right) \omega ^{\nu }=0,
\end{equation}
or, 
\begin{equation}\label{Proca1}
\nabla _{\mu }\tilde{F}^{\mu \nu }+\frac{3}{%
	2}M_p^2\alpha ^{2}\delta^2\mathcal{L}_m \omega ^{\nu }=0.
\end{equation}

In Riemann geometry the Weyl field strength $\tilde{F}^{\mu \nu }$
satisfies identically the equations
\begin{equation}\label{Proca2}
\nabla _{\sigma }\tilde{F}_{\mu \nu }+\nabla _{\mu }\tilde{F}_{\nu \sigma }+\nabla
_{\nu }\tilde{F}_{\sigma \mu }=0.
\end{equation}

By varying the action (\ref{Sf}) with respect to the metric tensor $g_{|mu\nu}$ we obtain the conformally invariant gravitational field equations with curvature-matter  coupling in the form
\begin{align}\label{feq}
&M_{p}^{2}\left[ \mathcal{L}_{m}R_{\mu \nu }+\left( g_{\mu \nu
}\nabla _{\alpha }\nabla ^{\alpha }-\nabla _{\mu }\nabla _{\nu }\right)
\mathcal{L}_{m}-\frac{3\alpha ^{2}}{2}\mathcal{L}_m\omega _{\mu }\omega _{\nu }\right]  \nonumber\\
&-\frac{1}{2}M_{p}^{2}\mathcal{T}_{\mu \nu }\left( R-\frac{3\alpha ^{2}}{2}%
\omega _{\alpha }\omega _{\beta }g^{\alpha \beta }+3\xi ^{2}M_{p}^{2}\right)
 +\frac{8}{%
	\eta ^{2}}B_{\mu \nu }-2\tilde{T}_{\mu \nu }^{(\omega)} =0,
\end{align}
where
\begin{equation}
\mathcal{T}_{\mu \nu }=g_{\mu \nu }+\frac{\xi ^{2}}{\gamma ^{2}}T_{\mu \nu },
\end{equation}%
\begin{equation}
\tilde{T}_{\mu \nu }^{(\omega)}=\frac{1}{2\delta ^{2}}\left( -\tilde{F}_{\mu
	\lambda }\tilde{F}_{\nu }^{~\lambda }+\frac{1}{4}\tilde{F}_{\lambda \sigma }%
\tilde{F}^{\lambda \sigma }g_{\mu \nu }\right),
\end{equation}
is the electromagnetic type energy-momentum tensor associated to the Weyl field, and $B_{\mu \nu }$, is the Bach tensor, given by
\begin{equation}
B_{\mu \nu }=\nabla _{\lambda }\nabla _{\sigma }C_{\mu~ \nu }^{~\sigma ~\lambda
}+\frac{1}{2}C_{\mu~ \nu }^{~\lambda ~\sigma }R_{\lambda \sigma }.
\end{equation}

Note that ${\cal T}_{\mu \nu}$ may be interpreted as an effective metric tensor that also functionally depends on the thermodynamic properties of matter. 
In four dimensions the Bach tensor is trace free, and also $\tilde{T}_{\mu}^{(\omega)\mu}=0$. Hence, by taking the trace of the field equations (\ref{feq}) we arrive at the scalar relation
\begin{eqnarray}\label{trace}
&&\left( \mathcal{L}_{m}R+3\nabla _{\alpha }\nabla ^{\alpha }\mathcal{L}_{m}-%
\frac{3\alpha ^{2}}{2}\mathcal{L}_m\omega ^{2}\right)
-\frac{1}{2}\mathcal{T}\left( R-%
\frac{3\alpha ^{2}}{2}\omega ^{2}+3\xi ^{2}M_{p}^{2}\right) =0,
\end{eqnarray}
where we have denoted $\omega ^{2}=\omega _{\mu }\omega ^{\mu }$. The trace equation can be also formulated as
  \begin{eqnarray}
&&\left( \mathcal{L}_{m}-\frac{1}{2}\mathcal{T}\right) R+3\nabla _{\alpha
}\nabla ^{\alpha }\mathcal{L}_{m}-\frac{3\alpha ^{2}}{2}\left( \mathcal{L}_m-\frac{1}{2}%
\mathcal{T}\right) \omega ^{2}
-\frac{3\xi ^{2}M_{p}^{2}}{2}\mathcal{T}=0.
\end{eqnarray}

By eliminating the term $\nabla _{\alpha }\nabla ^{\alpha }\mathcal{L}_{m}$ between Eq.~(\ref{feq}) and Eq.~(\ref{trace}) we find 
\begin{align}
&\frac{1}{2}M_{p}^{2}\left[ \mathcal{L}_{m}\left( R_{\mu \nu }-\frac{1}{3}%
g_{\mu \nu }R\right) -\frac{3\alpha ^{2}}{2}\mathcal{L}_m\left( \omega _{\mu }\omega
_{\nu }-\frac{1}{3}g_{\mu \nu }\omega ^{2}\right) \right]  \nonumber\\
&-\frac{1}{4}M_{p}^{2}\left( \mathcal{T}_{\mu \nu }-\frac{1}{3}g_{\mu \nu }%
\mathcal{T}\right) \left( R-\frac{3\alpha ^{2}}{2}\omega ^{2}+3\xi
^{2}M_{p}^{2}\right) 
\nonumber\\
&
-\frac{1}{2}M_{p}^{2}\nabla _{\mu }\nabla _{\nu }%
\mathcal{L}_{m} +\frac{4}{\eta ^{2}}B_{\mu \nu }
-\tilde{T}_{\mu \nu }^{(\omega)}=0.
\end{align}

With the help of the Einstein tensor the field equations can be written as
\begin{eqnarray}
&& R_{\mu \nu }-\frac{1}{2}g_{\mu \nu }R +\frac{8}{\eta
^{2}M_{p}^{2}\mathcal{L}_m}B_{\mu \nu }+\frac{1}{\mathcal{L}_{m}}\hat{\Sigma}_{\mu \nu }%
\mathcal{L}_{m}
+\frac{1}{2}\left( g_{\mu \nu }-\frac{\mathcal{T}_{\mu \nu }}{%
\mathcal{L}_{m}}\right) R\nonumber\\
&&=-\frac{3}{2}\frac{1}{\mathcal{L}_{m}}\left( \frac{\alpha ^{2}}{2}\omega
^{2}-\xi ^{2}M_{p}^{2}\right) \mathcal{T}_{\mu \nu }
+\frac{3\alpha ^{2}}{2}%
\omega _{\mu }\omega _{\nu }+\frac{2}{M_{p}^{2}}\frac{1}{\mathcal{L}_{m}}%
\tilde{T}_{\mu \nu }^{(\omega)},
\end{eqnarray}
where we have denoted $\hat{\Sigma}_{\mu \nu }=g_{\mu \nu }\nabla _{\alpha
}\nabla ^{\alpha }-\nabla _{\mu }\nabla _{\nu }$.

\subsection{Generalized Poisson equation, and corrections to the Newtonian potential}

By assuming that the coupling constant $\eta $ is large, we can neglect in the gravitational field equations the Bach tensor. Hence, Eqs.~(\ref{feq}) become 
\begin{eqnarray}\label{feqN}
R_{\nu }^{\mu } &=&-\frac{1}{\mathcal{L}_{m}}\left( \delta _{\nu }^{\mu
}\nabla _{\alpha }\nabla ^{\alpha }-\nabla ^{\mu }\nabla _{\nu }\right)
\mathcal{L}_{m}
+\frac{1}{2\mathcal{L}_{m}}\mathcal{T}_{\nu }^{\mu }\left( R-%
\frac{3\alpha ^{2}}{2}\omega ^{2}+3\xi ^{2}M_{p}^{2}\right) \nonumber\\
&&+\frac{3\alpha ^{2}}{2}\omega ^{\mu }\omega _{\nu }+\frac{2}{M_{p}^{2}%
\mathcal{L}_{m}}\tilde{T}_{\nu }^{(\omega )\mu }.
\end{eqnarray}

We investigate now the weak field and low velocity limit of Eqs.~(\ref{feqN}), by following the approach of \cite{LaLi}.
We assume that $\vec{v}^2 \ll 1$, and thus we can neglect the  spacelike components in $u^{\mu
}$, which in this limit has the components $u^{0}=u_{0}=1$, and
$u^{i }=0$, $i =1,2,3$, respectively. The matter energy-momentum tensor $T_{\nu
}^{\mu }=\rho u_{\nu }u^{\mu }$ has only one non-zero component, $%
T_{0}^{0}=\rho $, and in the weak field limit only the $g_{00}=1+2\varphi $ metric tensor component, where $\varphi $ is the Newtonian
gravitational potential, is different from the Minkowskian values of the metric \cite{LaLi}. In the same limit we have $R_{0}^{0}\approx R\approx \Delta \varphi$ \cite{LaLi}, and we can neglect in the gravitational field equations all the derivatives with
respect to the time.  Furthermore, we take $L_m=\rho $, $\mathcal{L}_{m}=1+\left(\xi^2/\gamma ^2\right)\rho $, and we assume that the time-like component of the Weyl vector field is dominant, so that $\omega ^2\approx \omega ^0\omega _0$. We also assume that the Weyl vector have a small spacelike variation, at least on the scale of the Solar System

Therefore, under these approximations, Eq.~(\ref{feqN}) gives
\begin{eqnarray}\label{genPo}
&&\left( 1+\frac{\xi ^{2}}{\gamma ^{2}}\rho \right) \Delta \varphi  =\frac{%
3\xi ^{2}}{\gamma ^{2}}\left( \frac{\alpha ^{2}}{2}\omega ^{2}+\xi
^{2}M_{p}^{2}\right) \rho 
+6\left( \xi ^{2}M_{p}^{2}-\frac{\alpha ^{2}}{2}%
\omega ^{2}\right) \varphi +
\frac{2\xi ^{2}}{\gamma ^{2}}\Delta \rho \nonumber\\
&&+3\left( \frac{\alpha ^{2}}{2}%
\omega ^{2}+\xi ^{2}M_{p}^{2}\right) .
\end{eqnarray}

Eq.~(\ref{genPo}) gives the generalized Poisson equation in the conformally invariant $f\left(R,L_m\right)$ theory.


If the curvature-matter coupling is weak, the constant $\gamma $ is very large. Therefore, if the matter density is low density, and in vacuum,  the generalized Poisson equation (\ref{genPo}) becomes
\begin{equation}
\Delta \varphi =6\left( \xi ^{2}M_{p}^{2}-\frac{\alpha ^{2}}{2}\omega
^{2}\right) \varphi +3\left( \frac{\alpha ^{2}}{2}\omega ^{2}+\xi
^{2}M_{p}^{2}\right).
\label{Poieq1}
\end{equation}

When the Weyl vector and the effective cosmological constant $\xi ^{2}M_{p}^{2}$ do vanish, we obtain the standard Newtonian vacuum Poisson
equation $\Delta \varphi =0$, which has the unique
spherically symmetric solution given by $\varphi (r)=-C/r$, where $C$ is an
integration constant. In the following for the sake of simplicity we neglect the term proportional to the potential in Eq.~(\ref{Poieq1}), by assuming  
\begin{equation}
\phi \ll \frac{\alpha ^2 \omega ^2/2+\xi ^2 M_p^2}{2\left(\xi ^2M_p^2-\alpha \omega ^2/2\right)}.
\end{equation}
holds.
 The above condition can be also formulated as
\begin{equation}\label{condnew}
r \gg r_g\left|\frac{\xi ^2M_p^2-\alpha \omega ^2/2}{\alpha ^2\omega ^2/2+\xi^2M_p^2}\right|,
\end{equation}
where we have assumed for the potential its Newtonian form $\phi (r)=\left|GM_{\odot}/r\right|$, and $r_g=2GM_{\odot}\approx 3\times 10^{5}$ cm, is the gravitational radius of the Sun. For $\alpha \omega ^2/2 \gg \xi ^2M_p^2$,  the approximation is valid for  $r \gg r_g$, that is, it works well in the standard Newtonian regime.

Therefore, Eq. (\ref{Poieq1}) takes the form
\begin{equation}
\frac{1}{r^{2}}\frac{d}{dr}\left( r^{2}\frac{d\varphi (r)}{dr}\right) =\frac{3
\alpha ^{2}}{2}\omega ^{2}(r)+3\xi ^{2}M_{p}^{2},
\end{equation}%
where we have assumed spherical symmetry. Its general solution is given by
\begin{equation}
\varphi (r)=-\frac{C}{r}+\frac{3\alpha ^{2}}{2}\int^{r}d\varsigma\frac{1}{\varsigma
^{2}}\int^{\varsigma }\theta ^{2}\omega ^{2}\left( \theta \right) d\theta +\frac{\xi
^{2}M_{p}^{2}}{2}r^{2}.
\end{equation}

By approximating $\omega ^{2}$ by a constant, an approximation that may be valid for some (astronomically) small regions of the spacetime, the gravitational potential is obtained as
\begin{equation}\label{potcorr}
\varphi (r)=-\frac{C}{r}+\frac{1}{2}\left(\frac{\alpha ^{2}\omega
^{2}}{2}+\xi ^{2}M_{p}^{2}\right) r^{2}.
\end{equation}

Therefore, the Weyl vector induces important modifications into the gravitational potential, and these corrections may lead to experimental or observational tests that may prove or disprove the presence of Weyl geometric effects  in the Universe.

\section{Summary and Conclusion}

In this proceedings, we considered modified theories of gravity with a linear coupling between matter and geometry. An interesting characteristic of these theories is the
non-conservation of the energy-momentum tensor, indicating
that matter and energy fluxes can be generated by the conversion of
curvature into matter.
While these theories offers an alternative explanation to the standard cosmological model for the expansion history of the universe, they offer a paradigm for nature fundamentally distinct from dark energy models of cosmic acceleration, even those that perfectly mimic the same expansion history.

 A fundamental concept in theoretical physics, initially discussed by Weyl \cite{Weyl1}, is the requirement of the conformal invariance of the physical laws. Conformal invariance is a highly attractive idea,  analogous to the gauge principle in elementary particle physics that played such an important role in the development of modern physics. The global transformations of units are similar to the global gauge transformations, or, equivalently, to global internal-symmetry transformations. Once we extend  the units transformations to the local level, and we require the conformal invariance of physical laws we obtain a situation comparable to the requirement  of gauge and internal invariance at the local level, a problem which is solved in elementary particle physics by the introduction of gauge fields. On the other hand, the fundamental laws of physics, including Maxwell's equations,  the electromagnetic, weak, and strong interactions, the massless scalar field equation, and the massless Dirac equation, respectively, are all conformally invariant. Hence, the world of microscopic physics is conformally invariant, while Einstein's gravity is not.

Therefore, there is one more fundamental difference between the microcosm of particle physics, and the macrocosm of the gravitational interaction. Abandoning the requirement of the conformal invariance of elementary particle physics is not an acceptable choice. But then a bridge between elementary particle physics and gravity can be built up by introducing the principle of conformal invariance in Einstein's general relativity. But this avenue would necessarily lead to the use of Weyl geometry to describe gravitational phenomena.

In the present paper we have presented the basic ideas and approach for one of the simplest conformally invariant models with curvature-matter coupling, with the field equations fully satisfying this requirement.  To construct a conformally invariant gravitational action in presence of matter {\it a coupling between matter and geometry is essentially necessary}. The simplest possible conformally invariant matter-curvature coupling is expressed by a term having the simple form $L_m\tilde{R}^2$, leading to a theory that is quadratic in the Weyl scalar $\tilde{R}$. However, by using the linear/scalar representation of the quadratic Weyl gravity \cite{Gh7, Gh4, Gh5, Gh6}, one can formulate the theory in the standard Riemannian geometry as a particular version of the $f\left(R,L_m\right)$ gravity theory \cite{Harko:2010mv}, in which the gravitational action is formulated in terms of an arbitrary analytic function of the (Riemannian) Ricci scalar $R$, and of the matter Lagrangian $L_m$. The theory that we have briefly reviewed in the present paper imposes a specific requirement on the general theory, namely, the condition of its conformal invariance at the level of the action, and of the field equations.

Conformally invariant Weyl type gravity theories can be also tested at cosmological scales, as done in \cite{GhHa}, where it was shown that they can represent attractive alternatives to the standard $\Lambda$CDM cosmological model.  
It is fundamental to understand how one may differentiate these modified theories of gravity from dark energy models
 (for instance, through structure formation).
Tests from the solar system, large scale structure and lensing essentially restrict the range of allowed modified gravity models.

Surveys such as the EUCLID space telescope, the Square Kilometre Array (SKA) radio telescope, the Dark Energy Survey (DES), and the Extended Baryon Oscillation Spectroscopic Survey (eBOSS) as part of the Sloan Digital Sky Survey III (SDSS)  will provide new opportunities to test the different cosmological models.
Indeed, with the wealth of unprecedented high precision observational data that will become available by these upcoming and planned surveys, we are dawning in a golden age of cosmology, which offers a window into understanding the perplexing nature of the cosmic acceleration, dark matter and of gravity itself.

\section{Acknowledgments}

FSNL acknowledges support from the Funda\c{c}\~{a}o para a Ci\^{e}ncia e a Tecnologia (FCT) Scientific Employment Stimulus contract with reference CEECINST/00032/2018, and the research grants No. UIDB/FIS/04434/2020, UIDP/FIS/04434/2020, No. PTDC/FIS-OUT/29048/2017, and No. CERN/FIS-PAR/0037/2019.
The work of TH is supported by a grant of the Romanian Ministry of Education and Research, CNCS-UEFISCDI, project number PN-III-P4-ID-PCE-2020-2255 (PNCDI III).

\bibliographystyle{ws-procs961x669}
\bibliography{ws-pro-sample}

\begin{thebibliography}{10}

\bibitem{SupernovaCosmologyProject:1998vns}
S.~Perlmutter \textit{et al.} [Supernova Cosmology Project],
``Measurements of $\Omega$ and $\Lambda$ from 42 high redshift supernovae,''
Astrophys. J. \textbf{517}, 565-586 (1999).

\bibitem{SupernovaSearchTeam:1998fmf}
A.~G.~Riess \textit{et al.} [Supernova Search Team],
``Observational evidence from supernovae for an accelerating universe and a cosmological constant,''
Astron. J. \textbf{116}, 1009-1038 (1998).



\bibitem{Nojiri:2006ri}
S.~Nojiri and S.~D.~Odintsov,
``Introduction to modified gravity and gravitational alternative for dark energy,''
eConf \textbf{C0602061}, 06 (2006)
[arXiv:hep-th/0601213].

\bibitem{Lobo:2008sg}
F.~S.~N.~Lobo,
``The Dark side of gravity: Modified theories of gravity,''
[arXiv:0807.1640 [gr-qc]].

\bibitem{Capozziello:2011et}
S.~Capozziello and M.~De Laurentis,
``Extended Theories of Gravity,''
Phys. Rept. \textbf{509}, 167-321 (2011)
[arXiv:1108.6266 [gr-qc]].

\bibitem{Nojiri:2010wj}
S.~Nojiri and S.~D.~Odintsov,
``Unified cosmic history in modified gravity: from F(R) theory to Lorentz non-invariant models,''
Phys. Rept. \textbf{505}, 59-144 (2011)
[arXiv:1011.0544 [gr-qc]].

\bibitem{Lobo:2014ara}
F.~S.~N.~Lobo,
``Beyond Einstein's General Relativity,''
J. Phys. Conf. Ser. \textbf{600}, no.1, 012006 (2015)
[arXiv:1412.0867 [gr-qc]].

\bibitem{Avelino:2016lpj}
P.~Avelino, T.~Barreiro, C.~S.~Carvalho, A.~da Silva, \textit{et al.}
``Unveiling the Dynamics of the Universe,''
Symmetry \textbf{8} (2016) no.8, 70
[arXiv:1607.02979 [astro-ph.CO]].



\bibitem{CANTATA:2021ktz}
E.~N.~Saridakis \textit{et al.} [CANTATA],
``Modified Gravity and Cosmology: An Update by the CANTATA Network,''
[arXiv:2105.12582 [gr-qc]].

\bibitem{Sotiriou:2008rp}
T.~P.~Sotiriou and V.~Faraoni,
``f(R) Theories Of Gravity,''
Rev. Mod. Phys. \textbf{82}, 451-497 (2010)
[arXiv:0805.1726 [gr-qc]].

\bibitem{Olmo:2011uz}
G.~J.~Olmo,
``Palatini Approach to Modified Gravity: f(R) Theories and Beyond,''
Int. J. Mod. Phys. D \textbf{20}, 413-462 (2011)
[arXiv:1101.3864 [gr-qc]].


\bibitem{Harko:2011nh}
T.~Harko, T.~S.~Koivisto, F.~S.~N.~Lobo and G.~J.~Olmo,
``Metric-Palatini gravity unifying local constraints and late-time cosmic acceleration,''
Phys. Rev. D \textbf{85}, 084016 (2012)
[arXiv:1110.1049 [gr-qc]].

\bibitem{Capozziello:2015lza}
S.~Capozziello, T.~Harko, T.~S.~Koivisto, F.~S.~N.~Lobo and G.~J.~Olmo,
``Hybrid metric-Palatini gravity,''
Universe \textbf{1}, no.2, 199-238 (2015)
[arXiv:1508.04641 [gr-qc]].

\bibitem{Harko:2018ayt}
T.~Harko and F.~S.~N.~Lobo,
{\it{Extensions of f(R) Gravity: Curvature-Matter Couplings and Hybrid Metric Palatini Theory}}, Cambridge University Press, Cambridge, (2018).

\bibitem{Bertolami:2007gv}
O.~Bertolami, C.~G.~Boehmer, T.~Harko and F.~S.~N.~Lobo,
``Extra force in f(R) modified theories of gravity,''
Phys. Rev. D \textbf{75}, 104016 (2007).


\bibitem{Harko:2010mv}
T.~Harko and F.~S.~N.~Lobo,
``f(R,$L_{m}$) gravity,''
Eur. Phys. J. C \textbf{70}, 373-379 (2010)
[arXiv:1008.4193 [gr-qc]].

\bibitem{Harko:2014gwa}
T.~Harko and F.~S.~N.~Lobo,
``Generalized curvature-matter couplings in modified gravity,''
Galaxies \textbf{2}, no.3, 410-465 (2014)
[arXiv:1407.2013 [gr-qc]].

\bibitem{Harko:2020ibn}
T.~Harko and F.~S.~N.~Lobo,
``Beyond Einstein\textquoteright{}s General Relativity: Hybrid metric-Palatini gravity and curvature-matter couplings,''
Int. J. Mod. Phys. D \textbf{29}, no.13, 2030008 (2020)
[arXiv:2007.15345 [gr-qc]].

\bibitem{Harko:2011kv}
T.~Harko, F.~S.~N.~Lobo, S.~Nojiri and S.~D.~Odintsov,
``$f(R,T)$ gravity,''
Phys. Rev. D \textbf{84}, 024020 (2011)
[arXiv:1104.2669 [gr-qc]].

\bibitem{Haghani:2013oma}
Z.~Haghani, T.~Harko, F.~S.~N.~Lobo, H.~R.~Sepangi and S.~Shahidi,
``Further matters in space-time geometry: $f(R,T,R_{\mu\nu}T^{\mu\nu})$ gravity,''
Phys. Rev. D \textbf{88},  044023 (2013).

\bibitem{Odintsov:2013iba}
S.~D.~Odintsov and D.~S\'aez-G\'omez,
``$f(R, T, R_{\mu\nu} T^{\mu\nu})$ gravity phenomenology and $\Lambda$CDM universe,''
Phys. Lett. B \textbf{725}, 437-444 (2013)
[arXiv:1304.5411 [gr-qc]].



\bibitem{Pri0} I. Prigogine and J. G\'{e}h\'{e}niau, ``Entropy, Matter, and Cosmology,'' Proc. Natl. Acad. Sci. USA \textbf{83}, 6245 (1986).

\bibitem{Pri} I. Prigogine, J. G\'{e}h\'{e}niau, E. Gunzig, and P. Nardone, ``Thermodynamics of Cosmological Matter Creation,'' Proc. Natl. Acad. Sci. USA \textbf{85}, 7428 (1988).


\bibitem{Weyl}  H. Weyl, ``Gravitation und Elektrizit\"{a}t,'' Sitzungsberichte der K\"{o}niglich Preussischen
Akademie der Wissenschaften zu Berlin {\bf 1918}, 465-478 (1918).

\bibitem{Weyl1} H. Weyl, Space, Time, Matter, Dover, New York, 1952

\bibitem{Scholz} E. Scholz, ``The unexpected resurgence of Weyl geometry in late 20th century physics,'' arXiv:1703.03187 (2017).

\bibitem{Bertolami:2008ab}
O.~Bertolami, F.~S.~N.~Lobo and J.~Paramos,
``Non-minimum coupling of perfect fluids to curvature,''
Phys. Rev. D \textbf{78}, 064036 (2008)
[arXiv:0806.4434 [gr-qc]].


\bibitem{Q1}
L. Parker, ``Particle creation in expanding universes,'' Phys. Rev. Lett. {\bf 21}, 562 (1968)

\bibitem{Q3}
L. Parker, ``Quantized fields and particle creation in expanding universes. 2,'' Phys. Rev. {\bf D3}, 2546 (1971).

\bibitem{Q5}
L. E. Parker and D. J.  Toms,
{\it{Quantum Field Theory in
Curved Spacetime-Quantized Fields and Gravity}};  Cambridge
University Press: Cambridge, UK, 2009.


\bibitem{Harko:2015pma}
T.~Harko, F.~S.~N.~Lobo, J.~P.~Mimoso and D.~Pav\'on,
``Gravitational induced particle production through a nonminimal curvature\textendash{}matter coupling,''
Eur. Phys. J. C \textbf{75}, 386 (2015)
[arXiv:1508.02511 [gr-qc]].

\bibitem{Gh7} D. M. Ghilencea, ``Standard Model in Weyl conformal geometry,'' arXiv:2104.15118 (2021).

\bibitem{LaLi} L. D. Landau and E. M. Lifshitz, The Classical Field Theory,
Pergamon Press, New York, 1975

\bibitem{Gh4} D. M. Ghilencea, ``Stueckelberg breaking of Weyl conformal geometry and applications to gravity,''
Phys. Rev. D {\bf 101}, 045010 (2020).

\bibitem{Gh5} D. M. Ghilencea, ``Palatini quadratic gravity: spontaneous breaking of gauged scale symmetry and inflation,'' Eur. Phys. J. C {\bf 80}, 1147 (2020).

\bibitem{Gh6} D. M. Ghilencea,  ``Gauging scale symmetry and inflation: Weyl versus Palatini gravity,'' Eur. Phys. J. C {\bf 81}, 510 (2021).

\bibitem{GhHa} D. M. Ghilencea and T. Harko, ``Cosmological evolution in Weyl conformal geometry,'' arXiv:2110.07056 (2021). 




\end{thebibliography}


\end{document}